\documentclass[prd,twocolumn,notitlepage,longbibliography,nofootinbib,preprintnumbers]{revtex4-1}
\usepackage[utf8]{inputenc}

\usepackage{bm} 
\usepackage{comment} 
\usepackage[colorlinks=true,urlcolor=black,anchorcolor=black,citecolor=black,linkcolor=black,filecolor=black,menucolor=black,pagecolor=black,linktocpage=true,pdfproducer=medialab,pdfa=true]{hyperref}
\usepackage{amsmath,latexsym,amssymb,mathrsfs}
\usepackage{graphicx}

\begin{document}
\title{2-form gauge theory dual to scalar-tensor theory}
\author{Daisuke Yoshida}
\email{dyoshida@hawk.kobe-u.ac.jp}
\affiliation{Department of Physics, Kobe University, Kobe 657-8501, Japan}
\preprint{KOBE-COSMO-19-08}

\begin{abstract}
We generalize the electromagnetic duality between a massless, canonical scalar field and a 2-form gauge field in 4-dimensional spacetime to scalar-tensor theories. We derive the action of 2-form gauge field that is dual to two kinds of scalar-tensor theories: shift symmetric K-essence theory and the shift symmetric Horndeski theory up to quadratic in scalar field. The former case, the dual 2-form has a nonlinear kinetic term. The latter case, the dual 2-form has non-trivial interactions with gravity through Einstein tensor. In both case, the duality relation is modified from usual case, that is, the dual 2-form field is not simply given by the Hodge dual of the gradient of the scalar field.  
\end{abstract}
\maketitle

\section{Introduction}
A scalar field interacting with gravity through non-trivial coupling enables us to construct diverge scenarios of both early and late-time Universe. As the first guideline to look for possible forms of interactions, it is important to classify interactions based on the presence/absence of Ostrogradsky's ghost instability (see \cite{Woodard:2015zca} for a review).
Understanding on ghost free scalar interactions has been deepened especially after the re-finding of Horndeski theory~\cite{Horndeski:1974wa} as a generalized Galileon theory \cite{Deffayet:2011gz,Kobayashi:2011nu} (see also Ref.\cite{Kobayashi:2019hrl} for a recent review), which is the most general scalar-tensor theory with second-order derivatives in the Euler-Lagrange equations. It was found that Horndeski theory is not the most general theory avoiding Ostrogradsky's ghost \cite{Gleyzes:2014dya,Gleyzes:2014qga}. It is possible to construct a theory which has higher derivative terms in the Euler-Lagrange equations apparently but these equations can be rewritten as a set of the equations up to second order time derivatives. At quadratic order in $\nabla_{\mu} \nabla_{\nu} \phi$, where $\phi$ is a scalar field and $\nabla_{\mu}$ is the covariant derivative, the general scalar tensor theory is found in Refs.~\cite{Langlois:2015cwa,Langlois:2015skt,Crisostomi:2016czh,Achour:2016rkg}, which is called extended scalar-tensor theory \cite{Crisostomi:2016czh} or degenerate higher order derivative scalar tensor (DHOST) theory~\cite{Achour:2016rkg}.  Cubic order version of DHOST theory is investigated in Ref.\cite{BenAchour:2016fzp}.

Not only interactions with a scalar field, but also that with other type of fields has been actively investigated.  For $U(1)$ gauge field, Horndeski also found the most general theory with second order Euler-Lagrange equations called vector-Horndeski theory \cite{Horndeski:1976gi}.
If one relax $U(1)$ gauge symmetry and consider massive vector field, non-trivial interaction with gravity has been developed by an analogy of Horndeski theory for scalar field. This is called generalized Proca theory \cite{Heisenberg:2014rta}. As similar to the scalar-tensor theory, generalized Proca theory has its further generalization without exciting extra degrees of freedom. The most general theory up to quadratic in $\nabla_{\mu} A_{\nu}$, where $A_{\mu}$ is a vector field, is called extended vector-tensor theory \cite{Kimura:2016rzw}, which is an analogy of the extended scalar-tensor (or DHOST) theory.
Even though structure of the extended vector-tensor theory is much similar to the extended scalar-tensor theory, vector theory has important property:  It has been shown that a non-trivial branch of the extended vector-tensor theory, that does not include generalized Proca theory, has stable cosmological solutions \cite{Kase:2018tsb}, though similar branch of DHOST theory does not \cite{deRham:2016wji}. 

Our interest here is interactions between gravity and 2-form gauge field because 2-form gauge field is an essential ingredient of low energy limit of string theory as well as scalar and 1-form vector field. Especially, in type-I and heterotic string theory, 2-form field has interactions through Chern-Simons term of gravity. Then it was shown that, in 4 dimensional-spacetime, such 2-form can be regarded as a canonical scalar field with Chern-Simons coupling, that is axion, through the electromagnetic duality \cite{Campbell:1990fu,Campbell:1992hc, Svrcek:2006yi}: As well known, there is a duality between $p$ and $D - p - 2 $ form field in $D$-dimension at least when there is no interaction. See Ref.\cite{Alexander:2009tp} for a review of Chern-Simons coupling to scalar field. The electromagnetic duality between scalar and 2-form field is useful even in the context of black hole hair.  It was discussed that trivial-black hole solution with vanishing scalar field can have ``axionic hair'' in the dual description \cite{Bowick:1988xh}. In the context of cosmology, 2-form field also has  much interest because it expected to produce statistical anisotropy \cite{Ohashi:2013mka,Ito:2015sxj,Ito:2015sxj,Obata:2018ilf,Almeida:2019xzt}.   

The purpose of this paper is to develop the way to extend such duality of canonical scalar field to a subclass of Horndeski theory. As well as the resultant 2-form theory dual theory will provide an new example of healthy interactions between 2-form field and gravity in 4-dimension, it might be a good candidate for healthy theory even in higher dimension as for the Chern-Simons case. Note that 2-form and scalar field duality in mimetic theory was discussed in \cite{Gorji:2018okn}.

This paper is organized as follows.
In the next section, we review the derivation of the electromagnetic duality between a canonical scalar field and a 2-form field with including Chern-Simons coupling. After that, we generalize this procedure to so called K-essence theory and the shift symmetric Horndeski theory up to quadratic in $\phi$ in section \ref{sec3} and \ref{sec4} respectively. In section \ref{sec5}, we briefly see that our method cannot apply directly to more general class of Horndeski theory. The final section is devoted by summary and discussions. 

\section{Duality between canonical scalar field and 2-form field}
Let us begin with reviewing well known duality between a massless scalar field with Chern-Simons coupling and a 2-form field \cite{Campbell:1990fu,Campbell:1992hc, Svrcek:2006yi}. Let us consider the following action \cite{Jackiw:2003pm},
\begin{align}
 S^{\phi} & = \int d^4 x \sqrt{-g} \left[
\frac{M_{pl}^2}{2}R - \frac{\alpha}{2} g^{\mu\nu} \partial_{\mu} \phi \partial_{\nu} \phi + \kappa \phi I^{CS}
\right],\label{SphiCS}
\end{align}
where $I^{SC}$ is Chern-Simon term for gravity given by
\begin{align}
I^{CS} = \frac{1}{4} R_{\mu\nu\rho\sigma} \tilde{R}^{\mu\nu\rho\sigma}.
\end{align}
Here $\tilde{R}_{\mu\nu\rho\sigma}$ are components of Hodge dual of Riemann 2-form,
\begin{align}
 \tilde{R}_{\mu\nu\rho\sigma} := \frac{1}{2}\epsilon_{\mu\nu}{}^{\alpha\beta}R_{\alpha\beta\rho\sigma}.
\end{align}
$\epsilon^{\mu\nu\rho\sigma}$ is Levi-Civita tensor associated with $g_{\mu\nu}$,
\begin{align}
 \epsilon^{\mu\nu\rho\sigma} = - \frac{1}{\sqrt{-g}} [\mu\nu\rho\sigma], \epsilon_{\mu\nu\rho\sigma} = \sqrt{-g} [\mu\nu\rho\sigma],
\end{align}
where $[\mu\nu\rho\sigma]$ is the complete antisymmetric tensor with $[0123] = 1$.
Here $\alpha$ is a dimensionless constant. Note that our analysis can apply even when $\alpha$ is a function of other field;  $\alpha = \alpha(\chi^{I})$ with some fields $\chi^{I}$.  $\kappa$ is a constant of mass dimension $- 1$. The simplest example of electromagnetic duality, that is duality between free scalar and free 2-form field, can be obtained by setting $\kappa = 0$ in the following analysis.
To treat a 2-form field, it would be useful to write the action by differential form language.
However, since our purpose is to extend duality relation to more general scalar-tensor theory and scalar-tensor theory is usually written in terms of component notations, we will examine electromagnetic duality in component notation. 

As well known, Chern-Simons term can be written as total derivative form,
\begin{align}
 I^{CS} = - \nabla_{\mu} J_{CS}^{\mu},
\end{align}
where $J_{CS}^{\mu}$ is the Chern-Simon current\footnote{
A concrete expression for $J_{CS}{}^{\mu}$ can be written in terms of spin connection $\omega^{a}{}_{b}$,
\begin{align*}
 J^{CS}_{\mu}dx^{\mu} = * \left( \omega^{a}{}_{b} \wedge d \omega^{b}{}_{a} + \frac{2}{3} \omega^{a}{}_{b} \wedge \omega^{b}{}_{c} \wedge \omega^{c}{}_{a} \right).
\end{align*}
Note that vierbein which define this spin connection is nothing to do with our coordinate basis $dx^{\mu}$ and one can freely choose the basis to represent the Chen-Simons current.
}.

Then after integrating by part, the action can be written as
\begin{align}
 S^{\phi} = \int d^4 x \sqrt{-g} \left[
\frac{M_{pl}^2}{2} R - \frac{\alpha}{2}g^{\mu\nu}\partial_{\mu}\phi \partial_{\nu} \phi + \kappa \partial_{\mu}\phi J^{\mu}_{CS}
\right].
\end{align}

A technically important step to derive the dual action is to introduce the intermediate action $S^{AB}$ which is defined as follows: 
\begin{align}
 S^{AB}  &= \int d^4 x \sqrt{-g} \Biggl[
\frac{M_{pl}^2}{2}R - \frac{\alpha}{2} A_{\mu} A^{\mu} + \kappa A_{\mu} J^{\mu}_{CS} \notag\\
&\qquad \qquad \qquad \qquad  + \frac{1}{2 } \epsilon^{\mu\nu\rho\sigma} B_{\mu\nu} \nabla_{\rho} A_{\sigma} 
\Biggr].
\end{align}
The first line is the original action $S^{\phi}$ where $\partial_{\mu} \phi$ is replaced with $A_{\mu}$.
$S^{AB}$ is equivalent with the original scalar action $S^{\phi}$ because the equation
\begin{align}
 \frac{1}{\sqrt{-g}}\frac{\delta S^{AB}}{\delta B^{\mu\nu}} = \frac{1}{2}\epsilon^{\mu\nu\rho\sigma} \nabla_{\rho} A_{\sigma}  = 0 \qquad ( \Leftrightarrow d A = 0)\label{dA=0}
\end{align}
guarantees the presence of a scalar potential for $A_{\mu}$. Thus Euler-Lagrange equation \eqref{dA=0} ensures that $A_{\mu}$ can be written as $A_{\mu} = \partial_{\mu} \phi$ at least locally and, by plugging this relation into $S^{AB}$, one can recover original action $S^{\phi}$.

The dual $2$-form action can be obtained by eliminating $A_{\mu}$ from the intermediate action $S^{AB}$.
$A_{\mu}$ can be eliminated by using the equation,
\begin{align}
\frac{1}{\sqrt{-g}} \frac{\delta S^{AB}}{\delta A_{\sigma}} & = - \alpha A^{\sigma} + \kappa J_{\mu}^{CS} - \frac{1}{2} \epsilon^{\mu\nu\rho\sigma} \nabla_{\rho} B_{\mu\nu}= 0,
\end{align}
which means
\begin{align}
 A &= \frac{1}{\alpha} \left( (*H) + \kappa J^{CS} \right) ,\label{CSA=} 
\end{align}
where we introduced the field strength 3-form $H$ of 2-form field $B$ which is given by $H = d B$ and $*$ represents the Hodge dual. In component notation, it can be expressed as 
\begin{align}
 H_{\mu\nu\rho} &= 3 \nabla_{[\mu} B_{\nu\rho]}, \qquad (*H)_{\mu} = \frac{1}{3!} \epsilon_{\mu}{}^{\nu\rho\sigma} H_{\nu\rho\sigma}.
\end{align}
By plugging above expression into $S^{AB}$, we obtain the action of 2-form dual to canonical free scalar field,
\begin{align}
 S^{B} &= \int d^4 x \sqrt{-g} \left[
\frac{M_{pl}^2}{2} R - \frac{1}{12 \alpha} \hat{H}_{\mu\nu\rho} \hat{H}^{\mu\nu\rho} \right],\label{SBCS}
\end{align}
where
\begin{align}
 \hat{H}_{\mu\nu\rho} = H_{\mu\nu\rho} + \kappa(* J^{CS})_{\mu\nu\rho}.
\end{align}
The action \eqref{SBCS} is nothing but 4-dimensional analogy of interaction between gravity and 2-form gauge field in type I or heterotic supergravity.
To summarize, we have examined the well known equivalence between scalar action $S^{\phi}$ \eqref{SphiCS} and 2-form gauge theory $S^{B}$ \eqref{SBCS}. The important step of derivation is to introduce the intermediate action $S^{AB}$. We will apply this method to more general scalar-tensor theory in the following sections.
Note that remembering the relation of $A$ to $\phi$, \eqref{CSA=} can be regarded as the duality relation between the original scalar field and 2-form gauge field, 
\begin{align}
 * d \phi = \frac{1}{\alpha}\hat{H} .
\end{align}
\section{Dual of shift symmetric K-essence theory}
\label{sec3}
\subsection{2-form dual action of K-essence theory}
Here we would like to extend above analysis to shift symmetric subclass of $K$-essence theory \cite{ArmendarizPicon:1999rj,Chiba:1999ka,ArmendarizPicon:2000dh} given by,
\begin{align}
 S^{\phi} = \int d^4x \sqrt{-g} \left[ \frac{M_{pl}^2}{2} R + K(X) \right],\label{SphiK}
\end{align}
where $X = g^{\mu\nu}\partial_{\mu} \phi \partial_{\nu} \phi$.

First we would like to introduce the intermediate action $S^{AB}$ by
\begin{align}
 S^{AB} = \int d^4x \sqrt{-g} \left[ \frac{M_{pl}^2}{2} R + K(Y) + \frac{1}{2} \epsilon^{\mu\nu\rho\sigma} B_{\mu\nu} \nabla_{\rho} A_{\sigma} \right],
\end{align}
where $Y = g^{\mu\nu} A_{\mu} A_{\nu}$.
The original action is recovered by plugging $A_{\mu} = \partial_{\mu} \phi$. This replacement is guaranteed by $d A = 0$, which can be derived from the $B_{\mu\nu}$ variation of $S^{AB}$.

Similar to the analysis in the previous section, the duality relation can be obtained from the $A_{\mu}$ variation of the intermediate action, 
\begin{align}
\frac{1}{\sqrt{-g}}\frac{\delta S^{AB}}{\delta A_\sigma} = 2 K'(Y) A^{\sigma} + \frac{1}{3 !} \epsilon^{\sigma\mu\nu\rho} H_{\mu\nu\rho} = 0. \label{eomAK}  
\end{align}
By using relation $A = d \phi$,  duality relation between original scalar and 2-form field can be understood as,
\begin{align}
- 2 K'(X) d \phi = * H,\label{dualK}
\end{align}
which now includes non-linear dependence of $X$;

In order to obtain 2-form action, we need to solve Eq. \eqref{eomAK} for $A_{\mu}$.
We can obtain equation written by $Y$ and $|H|^2 := H_{\mu\nu\rho} H^{\mu\nu\rho}$ by squaring Eq.\eqref{eomAK},
\begin{align}
 4 K'(Y)^2 Y = - \frac{1}{3 !} |H|^2.\label{KY=H}
\end{align}
By solving this equation for $Y$, $Y$ can be written as a function of $|H|^2$ implicitly; $Y = Y(|H|^2)$.\footnote{
Here we assume 
\begin{align*}
 (2 K''(Y) Y + K'(Y) ) K'(Y) \neq 0,
\end{align*}
so that \eqref{KY=H} has inverse at least locally.
Our assumption is not valid for cuscuton theory\cite{Afshordi:2006ad},  $K(X) \propto \sqrt{\pm X}$.
}

Now \eqref{eomAK} can be regarded as
\begin{align}
 A_{\sigma} = - \frac{1}{2K'(Y(|H|^2))} \frac{1}{3!} \epsilon_{\sigma}{}^{\mu\nu\rho} H_{\mu\nu\rho},\label{A=A(H)}
\end{align}
and one can safely eliminate $A^{\mu}$ from the intermediate action $S^{AB}$.
Finally we  obtain
\begin{align}
 S^{B} := S^{AB}|_{\eqref{A=A(H)}} = \int d^4x \sqrt{-g} \left[ \frac{M_{pl}^2}{2} R + F(|H|^2) \right],\label{SBK}
\end{align}
with
\begin{align}
F(|H|^2) =   K(Y) - 2 K'(Y) Y,\label{defF}
\end{align}
where $Y = Y(|H|^2)$ is a function of $|H|^2$ given by \eqref{KY=H}.
Thus the dual of $K$-essence theory is described by a 2-form field with non-linear kinetic term.
We would like to emphasize that our analysis hold even if $K$ depends on other field $\chi^{I}$, $K = K(\chi^{I}, X)$.

\subsection{Derivation of the original equations of motion}
For a complementary check, let us see the equivalence at the level of equations of motion.
First full set of equations of motion in the original system $S^{\phi}$ is given by
\begin{subequations} 
\begin{align}
\frac{2 M_{pl}^{-2}}{\sqrt{-g}} \frac{\delta S^{\phi}}{\delta g^{\mu\nu}} &= G_{\mu\nu} - \frac{2}{M_{pl}^2} T_{\mu\nu}^{\phi} = 0, \\
\frac{1}{\sqrt{-g}} \frac{\delta S^{\phi}}{\delta \phi} &=
- 2 \nabla^{\mu} (K'(X) \partial_{\mu} \phi) = 0,\label{KdSphidphi}
\end{align}
\end{subequations}
where the energy momentum tensor $T^{\phi}_{\mu\nu}$ is given by
\begin{align}
 T_{\mu\nu}^{\phi} = - K'(X) \partial_{\mu} \phi \partial_{\nu}\phi + \frac{1}{2} g_{\mu\nu} K(X).
\end{align}
We will check if we can derive same equations from dual action $S^{B}$.
The equations of motion from $2$-form action $S^{B}$ can be derived as,
\begin{subequations}
\begin{align}
\frac{2 M_{pl}^{-2}}{\sqrt{-g}} \frac{\delta S^{B}}{\delta g^{\mu\nu}} &= G_{\mu\nu} - \frac{2}{M_{pl}^2} T^{B}_{\mu\nu} = 0, \\
\frac{1}{\sqrt{-g}} \frac{\delta S^{B}}{\delta B_{\mu\nu}} &= - 6 \nabla^{\rho} ( F' H_{\rho}{}^ {\mu \nu} ) = 0,  \label{KdSBdB}
\end{align}
\end{subequations}
where $T^{B}_{\mu\nu}$ is now given by
\begin{align}
 T^{B}_{\mu\nu} = - 3 F' H_{\mu \rho\sigma} H_{\nu}{}^{\rho\sigma} + \frac{1}{2} g_{\mu\nu} F.\label{KTB}
\end{align}
First let us focus on the Eq. \eqref{KdSBdB}.
By introducing 3-form
\begin{align}
 F' H := \frac{1}{3 !} F'(|H|^2) H_{\mu\nu\rho} dx^{\mu} \wedge dx^{\nu} \wedge dx^{\rho}, 
\end{align}
it can be written as
\begin{align}
 * d * (F' H ) =  \frac{1}{2} \nabla^{\rho} ( F' H_{\rho}{}^ {\mu \nu} ) dx^{\mu} \wedge dx^{\nu} = 0.
\end{align}
Thus Eq.\eqref{KdSBdB} ensures that we can introduce scalar potential $\phi$ for $* F' H$,
\begin{align}
 * F' H = -\frac{1}{12} d \phi,\label{*FH=}
\end{align}
where a coefficient is determined so that above relation reproduce $H = * d \phi $ for free 2-form field $F = - |H|^2/12$.
Let us define a function $K(Y)$ and $Y(|H|^2)$ by \eqref{defF} and \eqref{KY=H}. Then we can show the relation,
\begin{align}
 F'(|H|^2) &= - (K'(Y) + 2 Y K''(Y) ) \frac{d Y}{d |H|^2}  \notag\\
&= - \frac{1}{4 K'(Y)} \frac{d}{d |H|^2} \left(4 K'(Y)^2 Y \right) \notag\\
&= \frac{1}{4 ! K'(Y)}. 
\end{align}
Then we can rewrite duality relation \eqref{*FH=} as
\begin{align}
 H &= - \frac{1}{12 F'(|H|^2)} * d \phi \notag \\
&= - 2 K'(Y(|H|^2)) * d \phi,
\end{align}
or in component notation, by
\begin{align}
 H_{\mu\nu\rho} &= - 2 K'  \epsilon_{\mu\nu\rho}{}^{\sigma} \partial_{\sigma} \phi.\label{H=KYdphi}
\end{align}
By squaring Eq. \eqref{H=KYdphi}, we obtain
\begin{align}
\frac{1}{3!} |H|^2 = - 4 (K'(Y(|H|^2)))^2 g^{\mu\nu} \partial_{\mu} \phi \partial_{\nu} \phi.
\end{align}
By comparing it with Eq. \eqref{KY=H}
Then we obtain
\begin{align}
Y(|H|^2)  = \partial_{\mu} \phi \partial_{\nu} \phi =:  X ,
\end{align}
and duality relation can be written as
\begin{align}
 H = - 2 K'(X) * d \phi,
\end{align}
which is equivalent with \eqref{dualK}.
The equations of motion for $\phi$ \eqref{KdSphidphi} can be obtained from Bianchi identity $d H = 0$ as
\begin{align}
 0 &= * d H =  * d  (- 2 K'(X) * d \phi) \notag \\
&= -2 \nabla^{\mu} \left(
K'(X) \nabla_{\mu} \phi \right).
\end{align}
Finally the equivalence of the energy-momentum tensor can be confirmed just by substituting \eqref{dualK} into \eqref{KTB} as follows:
\begin{align}
T^{B}_{\mu\nu} &= - 3 F' H_{\mu\rho\sigma} H_{\nu}{}^{\rho\sigma} + \frac{1}{2} g_{\mu\nu} F \notag\\
 &= - \frac{3}{12^2 F' } \epsilon_{\mu\rho\sigma}{}^{\alpha} \epsilon_{\nu}{}^{\rho\sigma\beta} \partial_{\alpha} \phi \partial_{\beta} \phi + \frac{1}{2} g_{\mu\nu} F \notag \\
 &= K' \left( g_{\mu\nu} X - \partial_{\mu}\phi \partial_{\nu}\phi \right) + \frac{1}{2} g_{\mu\nu} F \notag \\
&= - K' \partial_{\mu} \phi \partial_{\nu} \phi +\frac{1}{2} g_{\mu\nu} (2 K' X + F)  \notag \\
&= - K' \partial_{\mu} \phi \partial_{\nu} \phi +\frac{1}{2} g_{\mu\nu} K = T^{\phi}_{\mu\nu}. 
\end{align}

\section{Dual of a subclass of shift symmetric Horndeski theory}
\label{sec4}
\subsection{2-form dual action of Horndeski theory} 
Next let us includes non-trivial interaction through Einstein tensor,
\begin{align}
 S^{\phi} &= \int d^4 x \sqrt{-g} \left[
\frac{M_{pl}^2}{2} (R - 2 \Lambda) - \frac{1}{2} {\cal G}^{\mu\nu} \partial_{\mu}\phi \partial_{\nu} \phi
\right],\label{ScalG}
\end{align}
where effective metric ${\cal G}_{\mu\nu}$ is defined by 
\begin{align}
 {\cal G}_{\mu\nu} := \alpha  g_{\mu\nu} + \beta G_{\mu\nu} .
\end{align}
Here $\alpha$ is a dimension less constant and $\beta$ has mass dimension $-2$.
We note that following analysis can apply even when  $\alpha$ and $\beta$ depends on other fields. This action is free of ghost instability due to higher derivative because it is included by Horndeski theory.
The action of Horndeski theory is given by
\begin{align}
 S^{Horn} = \int d^4 x \sqrt{- g}\sum_{n=2}^4 {\cal L}^{Horn}_n,
\end{align}
where ${\cal L}_n$ are defined by
\begin{subequations}
\begin{align}
 {\cal L}^{Horn}_2 &= G_2(\phi, X), \\
 {\cal L}^{Horn}_3 &= G_3(\phi, X)\Box \phi ,\\
 {\cal L}^{Horn}_4 &= G_4(\phi, X)R - 2 G_{4,X}(X)( \Box \phi^2 - (\nabla \nabla \phi)^2 ),\\
 {\cal L}^{Horn}_5 &= G_5(\phi, X)G^{\mu\nu} \nabla_{\mu}\nabla_{\nu}\phi \notag\\
& + \frac{1}{3} G_{5,X} (\Box \phi^3 - 3 \Box \phi (\nabla\nabla \phi)^2 +2 (\nabla \nabla \phi)^3 ).
\end{align}
\end{subequations}
Then our action $S^{\phi}$ corresponds to the following choice of the arbitrary functions,
\begin{subequations}
\begin{align}
 G_2(\phi,X) &= M_{pl}^2 \Lambda - \frac{\alpha + c_1}{2} X, \\
 G_3(\phi,X) &=  c_1 \phi,\\
 G_4(\phi,X) &= \frac{M_{pl}^2}{2} - \frac{\beta+c_2}{2} X, \\
 G_5(\phi,X) &= c_2  \phi.
\end{align}
\end{subequations}
Here $c_1$ and $c_2$ describe redundancy of arbitrary functions and we can set $c_1 = c_2 = 0$ without loss of generality. 
Note also that \eqref{ScalG} is the most general subclass of Horndeski theory that contains terms up to quadratic in $\phi$ and has shift symmetry.

The intermediate action $S^{AB}$ can be obtained as
\begin{align}
 S^{AB} = \int d^4 x \sqrt{-g} &\Biggl[
\frac{M_{pl}^2}{2} (R - 2 \Lambda) - \frac{1}{2} {\cal G}^{\mu\nu}  A_{\mu} A_{\nu} \notag\\
&\qquad \qquad + \frac{1}{2} \epsilon^{\mu\nu\rho\sigma}B_{\mu\nu}\nabla_{\rho}A_{\sigma}
\Biggr].
\end{align}

To remove auxiliary field $A_{\mu}$, we vary $S^{AB}$ by $A_{\mu}$ and obtain equation
\begin{align}
\frac{1}{\sqrt{-g}} \frac{\delta S^{AB}}{\delta A_{\sigma}} &= - {\cal G}^{\sigma\mu} A_{\mu} - \frac{1}{2} \epsilon^{\mu\nu\rho\sigma} \nabla_{\rho}B_{\mu\nu} \notag\\
&= - {\cal G}^{\sigma\mu} A_{\mu} + \frac{1}{3!} \epsilon^{\sigma\mu\nu\rho} H_{\mu\nu\rho} = 0 .
\end{align}
Contrary to the case of K-essence theory, this can be easily solved for $A_{\mu}$, provided that ${\cal G}^{\mu\nu}$ has the inverse matrix,
\begin{align}
 A_{\alpha} &= \frac{1}{3!} {\cal G}^{-1}_{\alpha\sigma}  \epsilon^{\sigma\mu\nu\rho} H_{\mu\nu\rho} = {\cal G}^{-1}{}_{\alpha}{}^{\sigma} (* H)_{\sigma}. \label{HornAtoB}
\end{align}
Since $A$ can be understood as $ d \phi$, it means the duality relation between scalar and 2-form field depends on Einstein tensor $G_{\mu\nu}$,
\begin{align}
  d\phi &=  {\cal G}^{-1}{}_{\alpha}{}^{\sigma} (* H)_{\sigma} dx^{\alpha}. \label{dualHorn}
\end{align}

The dual $2$-form action $S^{B}$ can be obtained by eliminating $A_{\mu}$ from the intermediate action $S^{AB}$ by using Eq. \eqref{HornAtoB}:
\begin{align}
 S^{B} &:= S^{AB}|_{\eqref{HornAtoB}} \notag\\
& = \int d^4 x \sqrt{-g} \Biggl[
\frac{M_{pl}^2}{2} (R - 2 \Lambda) \notag\\
&\qquad \qquad + \frac{1}{2 \cdot 3!^2} {\cal G}^{-1}_{\mu\nu} \epsilon^{\mu\rho\sigma\tau} \epsilon^{\nu\alpha\beta\gamma} H_{\rho\sigma\tau} H_{\alpha\beta\gamma} 
\Biggr].
\end{align}
Thus, 2-form fields interact with gravity through not only metric but also Einstein tensor $G_{\mu\nu}$ as similar with the original scalar-tensor theory. We assumed that ${\cal G}_{\mu}{}^{\nu}$ has inverse matrix ${\cal G}^{-1}{}_{\mu}{}^{\nu}$, this condition generally depends on spacetime metric. For example, if we consider theory with $\alpha = 0$, then 2-form description can be defined only for the spacetime $R_{\mu\nu} \neq 0$.  

It is useful to derive the expression without inverse matrix ${\cal G}^{-1}$.
It can be achieved by using relation, 
\begin{align}
 {\cal G}^{-1}{}_{\mu}{}^{\nu} {\cal G}^{-1}{}_{\rho}{}^{\iota}{\cal G}^{-1}{}_{\sigma}{}^{\kappa}{\cal G}^{-1}{}_{\tau}{}^{\lambda} \epsilon^{\mu\rho\sigma\tau} \notag = \frac{1}{\det{({\cal G}_{\cdot}{}^{\cdot})}} \epsilon^{\nu\iota\kappa\lambda},
\end{align}
where $\det{{\cal G}_{\cdot}{}^{\cdot}}$ represents the determinant of a matrix ${\cal G}_{\mu}{}^{\nu}$.
Actually 2- form action can be evaluated as
\begin{align}
& {\cal G}^{-1}{}_{\mu}{}^{\nu} \epsilon^{\mu\rho\sigma\tau} \epsilon_{\nu\alpha\beta\gamma} H_{\rho\sigma\tau} H^{\alpha\beta\gamma}  \notag\\
& = \frac{1}{\det{({\cal G}_{\cdot}{}^{\cdot})}} \epsilon^{\nu\iota\kappa\lambda} \epsilon_{\nu\alpha\beta\gamma} {\cal G}_{\iota}{}^{\rho} {\cal G}_{\kappa}{}^{\sigma} {\cal G}_{\lambda}{}^{\tau} H_{\rho\sigma\tau} H^{\alpha\beta\gamma} \notag\\
& =  \frac{3 !}{\det{({\cal G}_{\cdot}{}^{\cdot})}}
 {\cal G}_{\alpha}{}^{\rho} {\cal G}_{\beta}{}^{\sigma} {\cal G}_{\gamma}{}^{\tau} H_{\rho\sigma\tau} H^{\alpha\beta\gamma}.
\end{align}
Finally we obtain 2-form action which is dual to \eqref{ScalG} as 
\begin{align}
 S^{B}
&= \int d^4 x \sqrt{-g} \Bigl[
\frac{M_{pl}^2}{2} (R - 2 \Lambda) \notag\\
& \qquad \qquad - \frac{1}{12} \frac{1}{\det({\cal G}_{\cdot}{}^{\cdot})} {\cal G}^{\rho}{}^{\alpha} {\cal G}^{\sigma}{}^{\beta} {\cal G}^{\lambda}{}^{\gamma} H_{\rho\sigma\lambda} H_{\alpha\beta\gamma}
\Bigr]. \label{SBH}
\end{align}
This is the main result of this paper. We found a ghost free non-trivial interaction between 2-form field and curvature of spacetime.
\subsection{Derivation of the original equations of motion}
Let us confirm the equivalence between \eqref{ScalG} and \eqref{SBH} at the level of equations of motion.
First, the original equations of motion from scalar-tensor theory can be evaluated as,
\begin{align}
 \frac{2 M_{pl}^{-2}}{\sqrt{-g}} \frac{\delta S^{\phi}}{\delta g^{\mu\nu}} &= G_{\mu\nu} + \Lambda g_{\mu\nu} - \frac{2}{M_{pl}^2} T^{\phi}_{\mu\nu} = 0, \\
 \frac{1}{\sqrt{-g}} \frac{\delta S^{\phi}}{\delta \phi} &= \nabla_{\mu} \left( {\cal G}^{\mu\nu} \nabla_{\nu} \phi \right) = 0.\label{eomphiH} 
\end{align}
Here $T_{\mu\nu}^{\phi}$ is given by
\begin{align}
 T_{\mu\nu}^{\phi} = \frac{1}{2} {\cal O}_{\mu\nu}{}^{\rho\sigma} \partial_{\rho} \phi \partial_{\sigma} \phi  + \frac{1}{2} g_{\mu\nu} \left(
- \frac{1}{2} {\cal G}^{\rho\sigma} \partial_{\rho} \phi \partial_{\sigma} \phi
\right),
\end{align}
where differential operator ${\cal O}$ is defined as
\begin{align}
({\cal O}_{\mu\nu}{}^{\rho\sigma}f_{\rho\sigma})(x)  
: = 
\frac{1}{\sqrt{-g(x)}}\int d^4 y \sqrt{-g(y)} f_{\rho\sigma}(y) \frac{\delta {\cal G}^{\rho\sigma}(y)}{\delta g^{\mu\nu}(x)},
\end{align}
and we do not need the concrete expression for ${\cal O}_{\mu\nu}^{\rho\sigma}$.

The equations of motion of this 2-form field is given by
\begin{align}
\frac{2 M_{pl}^{-2}}{\sqrt{-g}} \frac{\delta S^{B}}{\delta g^{\mu\nu}} &= G_{\mu\nu} + \Lambda g_{\mu\nu} - \frac{2}{M_{pl}^2} T^{B}_{\mu\nu} = 0, \\ 
 \frac{1}{\sqrt{-g}}\frac{\delta S^{B}}{\delta B_{\beta\gamma}}
& = - \frac{1}{3!} \nabla^{\alpha}\left( {\cal G}^{-1}{}_{\mu}{}^{\nu}  \epsilon^{\mu\rho\sigma\tau} \epsilon_{\nu \alpha \beta \gamma}  H_{\rho\sigma\tau} \right)  = 0, \label{HorndSBdB}
\end{align}
where $T_{\mu\nu}^{B}$ is now given by
\begin{align}
 T^{B}_{\kappa\lambda} &= \frac{1}{2\cdot 3 !^2} {\cal O}_{\kappa\lambda}{}^{\kappa' \lambda'} {\cal G}^{-1}_{\kappa' \mu} {\cal G}^{-1}_{\lambda' \nu} \epsilon^{\mu\rho\sigma\tau} \epsilon^{\nu\alpha\beta\gamma} H_{\rho\sigma\tau} H_{\alpha\beta\gamma} \notag \\
& + \frac{1}{2} g_{\kappa \lambda} \left[
- \frac{1}{2\cdot 3 !^2} {\cal G}^{-1}_{\mu\nu} \epsilon^{\mu\rho\sigma\tau} \epsilon^{\nu\alpha\beta\gamma} H_{\rho\sigma\tau} H_{\alpha\beta\gamma}
\right]. 
\end{align}

Let us define 1 form
\begin{align}
 {\cal G}^{-1} * H := \frac{1}{3!} {\cal G}^{-1}_{\mu}{}^{\nu} \epsilon_{\nu}{}^{\rho\sigma\lambda} H_{\rho\sigma\lambda} dx^{\mu}.
\end{align}
Then from the equation \eqref{HorndSBdB}, we obtain
\begin{align}
(* d ({\cal G}^{-1} * H))_{\beta\gamma}  & = - \frac{1}{3!} \nabla^{\alpha}\left( {\cal G}^{-1}{}_{\mu}{}^{\nu}  \epsilon^{\mu\rho\sigma\tau} \epsilon_{\nu \alpha \beta \gamma}  H_{\rho\sigma\tau} \right) = 0
\end{align}
That guarantees the presence of a scalar potential, 
\begin{align}
 {\cal G}^{-1} * H = d \phi,
\end{align}
which is equivalent with \eqref{dualHorn}.

Now it is clear that the Bianchi identity for $H$ reduces to the equations of motion \eqref{eomphiH},
\begin{align}
0 = * d H = \frac{1}{3 !} \epsilon^{\mu\nu\rho\sigma} \nabla_{\mu} H_{\nu\rho\sigma} = \nabla_{\mu}\left({\cal G}^{\mu\nu} \nabla_{\nu} \phi \right),
\end{align}
and energy momentum tensor of $B_{\mu\nu}$ coincides with that of $\phi$,
\begin{align}
 T^{B}_{\kappa\lambda} &= \frac{1}{2} {\cal O}_{\kappa\lambda}{}^{\kappa' \lambda'} {\cal G}^{-1}_{\kappa' \mu} {\cal G}^{-1}_{\lambda' \nu} \left( \frac{1}{3 !} \epsilon^{\mu\rho\sigma\tau} H_{\rho\sigma\tau} \right) \left( \frac{1}{3!} \epsilon^{\nu\alpha\beta\gamma} H_{\alpha\beta\gamma} \right)   \notag \\
& + \frac{1}{2} g_{\kappa \lambda} \left[
- \frac{1}{2} {\cal G}^{-1}_{\mu\nu} \left( \frac{1}{3!} \epsilon^{\mu\rho\sigma\tau} H_{\alpha\beta\gamma} \right) \left(  \frac{1}{3!}\epsilon^{\nu\alpha\beta\gamma} H_{\rho\sigma\tau} \right)
\right] \notag\\
&= \frac{1}{2} {\cal O}_{\kappa\lambda}{}^{\kappa' \lambda'}\partial_{\kappa'} \phi \partial_{\lambda'} \phi  + \frac{1}{2} g_{\kappa \lambda} \left[
- \frac{1}{2} {\cal G}^{\mu\nu} \partial_{\mu} \phi \partial_{\nu} \phi
\right] \notag\\
&= T^{\phi}_{\kappa\lambda}.
\end{align} 

\section{On further generalization}
\label{sec5}
We have derived 2-form dual of K-essence theory and shift symmetric Horndeski theory up to quadratic order in $\phi$. 
Before closing our discussion, let us investigate more general class of shift symmetric Horndeski theory,
\begin{align}
 G_{i}(\phi, X) := G_{i}(X).
\end{align}

Even for this class of theory, we can construct an intermediate action $S^{AB}$ by the analogy of previous discussion,
\begin{align}
 S^{AB} = \int d^4 x \sqrt{- g}
\left[
\sum_{n=2}^4 {\cal L}^{Horn}_n|_{\nabla \phi \rightarrow A} 
+ \frac{1}{2} \epsilon^{\mu\nu\rho\sigma} B_{\mu\nu} \nabla_{\rho} A_{\sigma}
\right]
.
\end{align}
 In the 2 examples that we have investigated in this paper, it was possible to eliminate $A_{\mu}$ from the intermediate action $S^{AB}$ by using the equation from $A$ variation. This fact enabled us to construct the dual 2-form action in these cases. However, this property does not hold for the general class of shift symmetric Horndeski theory because the equation of $A_{\mu}$ generally includes $\nabla_{\mu} A_{\nu}$.
Actually we can derive the equation 
\begin{align}
 &\frac{1}{\sqrt{-g}} \frac{\delta S^{AB}}{\delta A_\mu}
= 2 G_{2,Y} A^\mu + 2 G_{3,Y} \left( A^\mu \nabla_\nu A^\nu -  A_\nu \nabla^\mu A^\nu \right) \notag\\
&+ 2 G_{4,Y} A^\mu R - 4 G_{4,YY} A^\mu \left(
(\nabla_\nu A^\nu)^2 - (\nabla_\nu A_{\rho})^2
\right) + \cdots \notag\\
& + 2 G_{5,Y} \nabla_{\rho}A_{\nu} \left( A^\mu G^{\nu\rho}  -  A^\nu G^{\rho\mu} \right)  + \cdots 
\end{align}
Thus $\nabla_\mu A_\nu$ terms can be avoided when
\begin{align}
 G_{3,Y} = 0, G_{4,YY} = 0, G_{5,Y} = 0.\label{conditionforG}
\end{align}
Since when $G_3$ or $G_5$ is constant, the corresponding ${\cal L}_n$ term becomes total derivative, we can set $G_3 = G_5 = 0$. Then our discussion can be applied only for the subclass of the Horndeski theory with 
\begin{subequations} \label{genGs} 
\begin{align}
 G_2(X) &= K(X), \\
 G_3(X) &= 0,\\
 G_4(X) &= \frac{M_{\text{pl}}^2}{2} - \frac{\beta}{2} X,\\
 G_5(X) &= 0,
\end{align}
\end{subequations}
where we introduce arbitrary function $K$ and two integration constants of differential equations \eqref{conditionforG}, which are $M_{\text{pl}}^2$ and $\beta$. This results shows that 2-form dual action can be obtained only for the theory we have addressed in the previous two sections or combinations of these theories. Note that when both $K(X)$ and $G^{\mu\nu}\partial_{\mu} \phi \partial_{\nu} \phi $ exist at same time, it is generally difficult to solve equation 
\begin{align}
 2 K'(Y) A_{\mu} - \beta G_{\mu}{}^{\nu} A_{\nu} + (* H)_{\mu} = 0,
\end{align}
for $A_{\mu}$ explicitly.

We would like to emphasize that this result does not mean there is no 2-form dual of scalar-tensor theory beyond \eqref{genGs}. 
Actually we can consider a coupling through Gauss-Bonnet term,
\begin{align}
 S^{GB} = \int d^4 \sqrt{-g} \lambda \phi I^{GB},
\end{align}
where $I^{GB}$ is the Gauss-Bonnet invariant given by
\begin{align}
I^{GB} =   - \frac{1}{4} \tilde{R}_{\mu\nu\rho\sigma} \tilde{R}^{\mu\nu\rho\sigma} 
= R^2 - 4 R_{\mu\nu} R^{\mu\nu} + R_{\mu\nu\rho\sigma} R^{\mu\nu\rho\sigma}.
\end{align}
Since integration of $I^{GB}$ becomes surface term in 4 dimension, this coupling respects shift symmetry $\phi \rightarrow \phi + c$.  It was shown that the Gauss-Bonnet coupling corresponds to shift symmetric Horndeski theory with the choice of a function \cite{Kobayashi:2011nu,Charmousis:2011ea}
\begin{align}
 G_5 = - 4 \lambda \log |X|.
\end{align}
Clearly this is not included by \eqref{genGs}. Nonetheless, we can construct 2-form dual because there is Gauss-Bonnet current
\begin{align}
I^{GB} = - \nabla_{\mu} J^{\mu}_{GB},
\end{align}
where a concrete expression can be written by connection 1-form $\omega_{a}{}^{b}$ with respect to some vierbein as \cite{Yale:2010jy},
\begin{align}
 J^{GB}_{\mu} dx^{\mu} = (\omega^{ab} \wedge d \omega^{cd} + \frac{2}{3} \omega^{ab} \wedge \omega^{cf} \wedge \omega_{f}{}^{d}) \epsilon_{abcd}.
\end{align} 
Since the derivation of 2-form dual action for Chern-Simons coupling dose not need the detail of $J^{CS}_{\mu}$, one can also derive 2-form dual action with Gauss-Bonnet coupling by the same manner.  

To summarize, our method can directly apply to, in principle, the following action of scalar-tensor theory
\begin{align}
 S^{\phi} =& \int d^4 x \sqrt{-g}\Biggl[
 \frac{M_{pl}^2}{2} R + K(X) - \frac{\beta}{2} G^{\mu\nu} \partial_{\mu} \phi \partial_{\nu} \phi \notag\\ 
& \qquad \qquad \qquad  + \kappa \phi I^{CS} + \lambda \phi I^{GB}
\Biggr],\label{finaction}
\end{align}
though it might be difficult to write down a dual 2-form action analytically because of the non-linear dependence of the kinetic term $K(X)$. Note that here we include Chern-Simons interactions though it is not included by Horndeski theory.
In a presence of other fields $\chi^{I}$, our analysis can directly apply even when $K = K(\chi^{I}, X)$ and $\beta = \beta(\chi^{I})$.

\section{Summary and Discussion}
We derived 2-form theories which are dual to shift symmetric scalar-tensor theories included by Horndeski theory.
We explicitly showed the equivalence between \eqref{SphiK} and \eqref{SBK}, and between \eqref{ScalG} and \eqref{SBH} at the level of both action and equations of motion. 2-form field has non-linear kinetic term in \eqref{SBK} and has couplings through Einstein tensor in \eqref{SBH}. In both case, duality relations are modified from standard relation $d \phi = * d B$, as obtained in \eqref{dualK} and \eqref{dualHorn}. These actions provide non-trivial example of interaction between 2-form field and gravity.  We found that direct application of our method to general shift symmetric scalar-tensor theory beyond \eqref{finaction} does not work well.

Since our new interaction through Einstein tensor is equivalent to Horndeski theory, it definitely free of Ostrogradsky's instability in 4-dimension. Though we found that it is difficult to generalize our analysis to more general class of shift symmetric Horndeski theory, it might be possible to construct dual 2-form theory in the framework of DHOST theory. Or  more generally, it would be interesting to study the most general ghost free interaction between 2-form and gravity, which might include 2-form theory which does not have dual scalar description. Our duality hold only for 4 dimensional spacetime. Then it is also interesting to ask whether the ghost freeness holds in arbitrary dimension. 

Our new 2-form interaction would be interesting for applications to inflation and black hole physics.  For inflation, 2-form field which coupled with inflaton possibly produces a statistical anisotropy~\cite{Ohashi:2013mka,Ito:2015sxj,Ito:2015sxj,Obata:2018ilf,Almeida:2019xzt}. It is not clear how much our new interaction affects to it. For black hole physics, scalar-haired black hole solution of shift symmetric Horndeski theory \eqref{ScalG} was studied in Ref. \cite{Babichev:2013cya}. Then it is interesting to clarify the relation of this scalar hair to the axionic hair of black hole known in free field~\cite{Bowick:1988xh}.

\begin{acknowledgments}
D.Y. is supported by the JSPS Postdoctoral Fellowships No.201900294 .  D.Y. would like to thank Jiro Soda, Toshifumi Noumi, Kazufumi Takahashi, Ryo Namba and Tokiro Numasawa for fruitful discussion. D.Y. is grateful to YITP because discussion in the workshop YITP-W-18-15 ``3rd work shop on gravity and cosmology by young researchers'' was helpful to complete this work.  
\end{acknowledgments}

\bibliography{ref}
\end{document}